\documentclass{article}
\usepackage[a4paper, total={5.9in, 8.8in}]{geometry}
\usepackage[utf8]{inputenc}
\usepackage{amsmath}
\usepackage{upgreek}
\DeclareMathOperator*{\argmin}{arg\,min}
\usepackage{booktabs}
\usepackage{graphicx}
\usepackage{caption}
\usepackage{subcaption}
\usepackage{bm}
\usepackage[colorlinks=true,linkcolor=black,anchorcolor=black,citecolor=black,filecolor=black,menucolor=black,runcolor=black,urlcolor=black]{hyperref}
\usepackage{abstract}

\usepackage{authblk}

\usepackage{enumitem}
\usepackage{placeins} 
\usepackage{gensymb}
\newcommand{\supplementarysection}{%
  \setcounter{figure}{0}
  \let\oldthefigure\thefigure
  \renewcommand{\thefigure}{SI-\oldthefigure}
  \let\oldthetable\thetable
  \renewcommand{\thetable}{SI-\oldthetable}
  \begin{centering}\section*{Supplementary Information}\end{centering}
  \let\oldchapter\chapter
}
\usepackage[citestyle=ieee,firstinits=true,backend=biber]{biblatex}
\renewbibmacro{in:}{}

\title{Milliwatt-level UV generation using sidewall poled lithium niobate}
 
\author[1,2,*]{C.A.A. Franken}
\author[1,3,*]{S.S. Ghosh}
\author[1,4]{C.C. Rodrigues}
\author[1]{J. Yang}
\author[1]{C.J. Xin}
\author[1]{S. Lu}
\author[1]{D. Witt}
\author[1]{G. Joe}
\author[4]{G.S. Wiederhecker}
\author[2]{K.-J. Boller}
\author[1,$^\dagger$]{M. Lončar}

\affil[1]{School of Engineering and Applied Sciences, Harvard University, USA}
\affil[2]{Department of Science and Technology, University of Twente, NL}
\affil[3]{Department of Physics, Harvard University, USA}
\affil[4]{Gleb Wataghin Physics Institute, University of Campinas, BR}
\affil[*]{These authors contributed equally to this work.}
\affil[$\dagger$]{Electronic mail: loncar@seas.harvard.edu.}

\date{March 20, 2025}

\begin{document}
\maketitle

\vspace{-0.25cm}
\begin{abstract} 
\noindent \textbf{Integrated coherent sources of ultra-violet (UV) light are essential for a wide range of applications, from ion-based quantum computing and optical clocks to gas sensing and microscopy. Conventional approaches that rely on UV gain materials face limitations in terms of wavelength versatility; in response frequency upconversion approaches that leverage various optical nonlinearities have received considerable attention. Among these, the integrated thin-film lithium niobate (TFLN) photonic platform shows particular promise owing to lithium niobate's transparency into the UV range, its strong second order nonlinearity, and high optical confinement. However, to date, the high propagation losses and lack of reliable techniques for consistent poling of cm-long waveguides with small poling periods have severely limited the utility of this platform. Here we present a sidewall poled lithium niobate (SPLN) waveguide approach that overcomes these obstacles and results in a more than two orders of magnitude increase in generated UV power compared to the state-of-the-art. Our UV SPLN waveguides feature record-low propagation losses of 2.3 dB/cm, complete domain inversion of the waveguide cross-section, and an optimum 50\% duty cycle, resulting in a record-high normalized conversion efficiency of 5050 \%$\bm{\mathrm{W^{-1}cm^{-2}}}$, and 4.2 mW of generated on-chip power at 390 nm wavelength. This advancement makes the TFLN photonic platform a viable option for high-quality on-chip UV generation, benefiting emerging applications.} 
\end{abstract}
\vspace{0.25cm}

\noindent A wide range of emerging technologies, including ion trap-based quantum computers \cite{Moody2022}, optical clocks \cite{Ludlow2015}, super-resolved structured illumination microscopy \cite{Lin2022} and spectroscopy \cite{Wiederkehr2014} rely on efficient generation and delivery of coherent optical signals at ultra-violet (UV) wavelengths. Unfortunately, semiconductor laser diodes in this wavelength range are either not readily available or do not meet the tunability and linewidth requirements \cite{Bruzewicz2019c,Hasan2021,Yang2024}. To meet the demands of such applications, novel hybrid and heterogeneously integrated laser sources have been developed in the visible violet range \cite{Franken2023, Wunderer2023b}. However, extension into the UV remains difficult due to the lack of suitable semiconductor amplifiers.
\par Alternatively, nonlinear conversion based on bulk crystals like $\beta$-barium borate \cite{Poberaj2009} and lithium tantalate \cite{Meyn1997a} has been leveraged to realize robust, wavelength agile UV sources. Lithium niobate (LN) is another promising candidate, with a high optical nonlinearity and comparably large transparency window down to 350 nm \cite{Zhu2021a}, which has facilitated efficient frequency conversion using bulk LN periodically poled waveguides \cite{Sugita2001, Mizuuchi2003a}. UV generation has been demonstrated in bulk and thin-film LN using resonantly-enhanced second harmonic generation (SHG) in nanospheres \cite{Wang2021}, Cherenkov phase matching at an LN-BBO interface \cite{Guan2023}, and supercontinuum generation \cite{Cheng2024,Ludwig2024}, albeit at low powers.

Recently, the ultra-low loss thin-film lithium niobate (TFLN) photonic platform has emerged as a powerful approach to realize high performance electro-optic (EO) devices and chip-scale systems at infra-red and visible wavelengths \cite{Renaud2023,Wang2018}. In addition frequency comb sources spanning UV-to-visible ranges have been demonstrated in TFLN \cite{Wu2024}. State-of-the-art periodically poled TFLN waveguides have been used to produce up to 30 $\upmu$W of on-chip UV power through second harmonic generation \cite{Hwang2023}. However, this output is significantly lower than what is theoretically achievable, given the material’s high nonlinearity and the enhanced interaction enabled by the tight confinement of TFLN waveguides. The lower than expected efficiency of this implementation has been attributed to high propagation losses (7.6 dB/cm), strong deviation from the optimum duty cycle (90\% instead of ideal 50\%), and film thickness variations throughout the waveguide. Furthermore, the aforementioned poling method can show a high sensitivity of domain inversion to poling field variations \cite{Zhao2020thesis,Nagy2020}. Addressing all these issues while achieving the small $\upmu$m-order poling period throughout a cm-scale waveguides are among the greatest challenges for generating UV in TFLN at mW-level powers. Only in that power regime do such sources become practically relevant for applications such as quantum computing \cite{Bruzewicz2019c}.

Besides lowering propagation losses, recent progress on the standard \textit{pole-before-etch} method, as used by Hwang et al. \cite{Hwang2023}, can improve UV conversion efficiencies. Chen et al. report that adapting the poling period to account for film thickness variation greatly increases the efficiency for infra-red to visible upconversion \cite{Chen2023}. However, this method still does not account for the fabricated geometry, i.e., waveguide shape and etch depth variations. More recent work addresses this by poling the film after etching the waveguide and taking such variations into account \cite{Xin2024}. The drawback of this \textit{pole-after-etch} approach is the placement of the poling fingers on the etched part of the film (slab), only inverting the slab part of the waveguide cross-section. Since a large fraction of the optical mode resides in the unetched part of the waveguide (ridge), this partial inversion can reduce the efficiency to as low as 20\% of the maximum achievable efficiency in a completely inverted film \cite{Xin2024}. The partial inversion can be leveraged to facilitate inter-modal phase matching, but remains below adapted poling efficiencies \cite{Shi2024}. Adjusting the placement of the poling fingers in \textit{pole-after-etch} to achieve complete inversion has not been explored. However, earlier work at infra-red wavelengths has placed electrodes closer to the optical mode to improve local control over poling in bulk-LN \cite{Gui2009}, or to avoid the propagation loss that can occur in pole-before-etch waveguides due to differential etching of poled domains \cite{Mu2021}, but efficiencies have remained low, not exceeding 500\%$\mathrm{W^{-1}cm^{-2}}$.

\begin{figure}[t!]
   \centering
    \makebox[\textwidth][c]{\includegraphics[width=180mm]{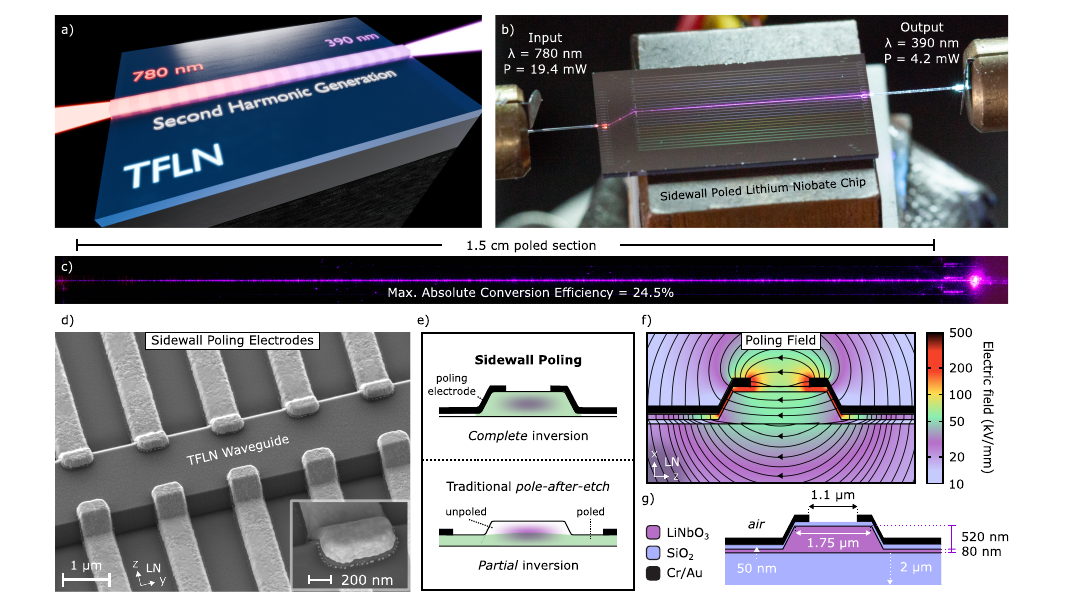}}%
   \caption{\textbf{UV generation using sidewall poled lithium niobate (SPLN) waveguides. } \textit{a) A second harmonic generation (SHG) process is used to upconvert visible 780 nm light to the ultra-violet (UV) at 390 nm. The quasi-phase matched nonlinear process is enabled by periodically inverting the sign of the $\chi^{(2)}$ nonlinearity in the thin-film lithium niobate (TFLN) waveguide using our sidewall poling method. b) Image of the sidewall poled lithium niobate chip recorded during UV generation. Specified powers indicate on-chip values. c) Top-down microscope image of the 1.5-cm long SPLN waveguide; here the UV light is being generated as the pump propagates through the waveguide from left to right. d) Scanning electron microscope image of metal electrodes used for sidewall poling. The electrodes reach from the TFLN slab (etched portion of the film) onto the waveguide ridge. By placing the poling fingers onto the waveguide, the duty cycle can be precisely controlled over long waveguide sections while ensuring that the entire waveguide cross-section is poled, thus realizing efficient conversion to the UV. Once established, the domain inversion remains stable, allowing for the removal of metal electrodes to minimize optical losses. e) Schematic showing a benefit of sidewall poling vs. traditional pole-after-etch methods; namely, complete inversion of the waveguide which can increase the normalized conversion efficiency significantly. 
   f) Simulation of the electric field pattern resulting from 127 V potential difference applied to the poling electrodes. Note that the field lines are nearly parallel to the crystalline z-axis, and, more importantly, the field strength in the entire waveguide cross-section is beyond the coercive field strength needed for domain inversion in thin-film lithium niobate ($\sim30$ kV/mm) \cite{Nagy2019}. g) Materials and dimensions for the waveguides and sidewall poling electrodes discussed in this work.}}
    \label{fig:1}
\end{figure}

Here we present an improvement by two orders of magnitude over the state of the art in UV power \cite{Hwang2023}, reaching the mW-level for the first time, using sidewall poled lithium niobate (SPLN) waveguides and second harmonic generation (Fig. \ref{fig:1}). Unique to our novel \textit{pole-after-etch} approach is the precise placement of the poling electrodes onto the sidewalls of the TFLN waveguide in combination with adapted poling \cite{Xin2024,Chen2023}. This allows us to take fabrication induced variations to the waveguide geometry into account while ensuring that each poled region is fully inverted across the entire waveguide cross-section, thus overcoming a major drawback of the pole-after-etch approach \cite{Xin2024,Su2024}. Using poling electrodes positioned on the waveguide sidewalls to a precision of 50 nm, we achieve a near-optimal (i.e., 50\%) duty cycle along 1.5-cm long waveguides and a reproducible robustness to poling field variations. These factors combined with low-loss (2.3 dB/cm) high confinement waveguides result in a record-high nonlinear conversion efficiency of 5050 \%$\mathrm{W^{-1}cm^{-2}}$ with a maximum 4.2 mW of generated on-chip UV power. 

\subsection*{Design and poling process}
\begin{figure}[t!]
    \centering
    \makebox[\textwidth][c]{\includegraphics[width=180mm]{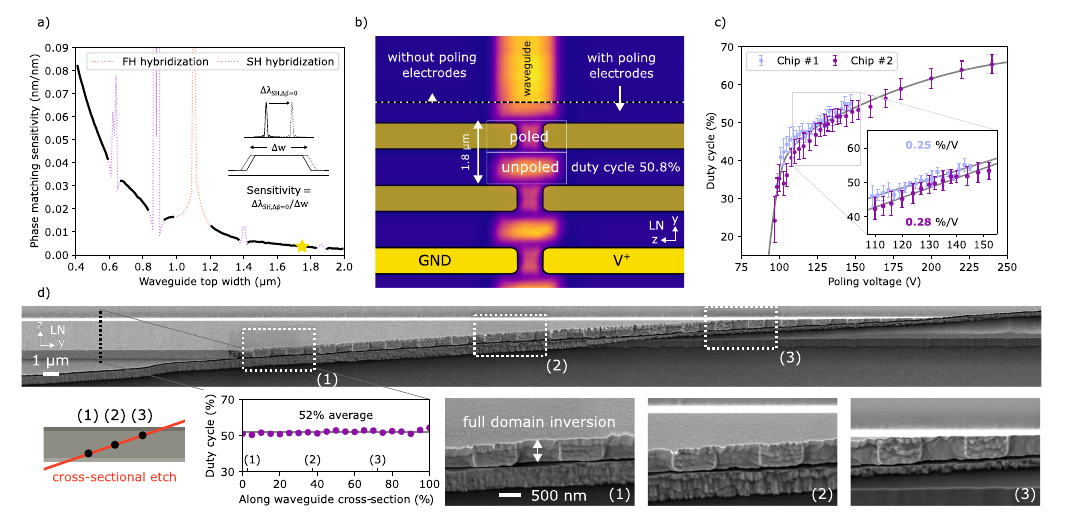}}%
    \caption{ \textbf{Optimization of the poling process. }\textit{a) Phase matching sensitivity, defined as the derivative of the optimally phase matched second harmonic wavelength with respect to waveguide top width, for different waveguide top widths. In other words, the sensitivity captures the change in the phase matched SH wavelength (at $\Delta \beta = 0$) as a function of variation in waveguide width. It can be seen that the phase matching condition for a wide waveguide is more resilient to fabrication tolerances that result in width variation. Our waveguide width of 1.75 $\mu$m (indicated by a star) minimizes phase matching sensitivity, to near zero, while avoiding divergences in the sensitivity originating from transverse mode hybridization (TE polarization fraction $<$90\%). b) Second harmonic microscope image showing sidewall poled domains in the waveguide ridge. The boundaries between inverted domains appear as dark horizontal stripes in the image. The poling electrodes are indicated in yellow. A near-optimum duty cycle is verified at 50.8\%. 
    c) Dependence of duty cycle on applied poling voltage for two fabricated chips. Each data point corresponds to a single test structure that is poled at a different poling voltage. At lower voltages below 100V the duty cycle falls off to the threshold voltage and at higher voltages the duty cycle saturates. The solid line based on a spline interpolation is provided as a visual reference. d) Scanning electron microscope images show that a diagonal cross-sectional cut through the waveguide that has been exposed to a differential wet etch reveals the domain inversion extending completely through the waveguide cross-section, from top to bottom. The duty cycle along the diagonally etched waveguide cross-section is inferred from the SEM image and plotted. A constant duty cycle indicates parallel domain walls, ensuring a constant phase matching condition across the entire mode.}}
    \label{fig:2}
\end{figure}

Achieving a device design with high conversion efficiency starts by determining the optimal waveguide geometry (Fig. \ref{fig:1}g). Recent work has highlighted that film thickness variations can negatively impact the phase matching conditions and efficiency \cite{Zhao2023,Xin2024}. Therefore, it is important to consider waveguide geometries that minimize the sensitivity of the phase matching condition to film thickness variation, such as waveguides in thicker films \cite{Kuo2021, Hwang2023}. Besides lower sensitivity, thicker waveguides have higher confinement and lower sidewall scattering loss, which can ultimately result in a higher conversion efficiency. Both considerations motivate our choice of a 600 nm thick lithium niobate film. Our choice of etch depth, around 500 nm, ensures full penetration of the poling field into the waveguide when using sidewall electrodes. In our work, any remaining sensitivity to film thickness and etch depth variations is mitigated by using adapted poling \cite{Zhao2023,Xin2024}. As a result, waveguide width variation becomes the dominant factor impacting phase matching \cite{Kuo2021}. Therefore, we investigate phase matching sensitivity to waveguide width variations in order to determine an optimal waveguide geometry. In Fig. \ref{fig:2}a we plot phase matching sensitivity, here defined as the derivative of the optimally phase matched second harmonic wavelength with respect to waveguide top width, for a range of waveguide widths (see Methods). The resulting curve shows that the phase matching in wider waveguides is more robust to width variation, from fabrication error for example. It can be seen that the chosen waveguide width of 1.75 $\upmu$m minimizes phase matching sensitivity, to near zero, while avoiding divergences in the sensitivity originating from transverse mode hybridization. 

Compared to traditional \textit{pole-after-etch} methods, sidewall poling positions electrodes directly from the slab onto the waveguide sidewall (Fig. \ref{fig:1}d and e), yielding several benefits. Simulation of poling electrodes on the waveguide sidewalls shows a uniform electric poling field across the entire waveguide which indicates homogeneous poling throughout the waveguide cross section (Fig. \ref{fig:1}f, calculated using a finite element mode solver), thereby realizing the complete cross section domain inversion which maximizes conversion efficiency (Fig. \ref{fig:1}e). The 1.75 $\upmu$m waveguide width also tightly confines the fundamental and second harmonic modes in the waveguide ridge, reducing sidewall scattering significantly (see Supplementary Information) while maintaining a high overlap between the interacting modes (98\%) for efficient conversion. While this waveguide width allows propagation of higher-order transverse modes at 390 nm, only the fundamental transverse mode at 780 nm is injected and the poling is designed to only phase match the fundamental TE-polarized modes at 780 and 390 nm. The fabrication process of the waveguide and poling electrodes is described in the Methods.

To ensure an optimum poling duty cycle (50\%), poling is undertaken iteratively, alternating between applying electrical pulses to poling electrodes and imaging the poled structures using a high resolution second harmonic microscope (details in Methods). Such images show dark lines between the inverted domains (Fig. \ref{fig:2}b), which allows for a straightforward extraction of the poling duty cycle (see Supplementary Information). The duty cycle as a function of poling voltage is characterized on two separately fabricated chips with the same design (Fig. \ref{fig:2}c). This measurement identifies an optimal poling voltage around 127 V, and shows that the duty cycle is robust to voltage variations: the duty cycle remains within 40 - 60\% when the poling voltage is within $\pm 20$\% of the optimal voltage (Fig. \ref{fig:2}c inset). The poled duty cycle as a function of voltage also shows nearly the same slope in the 50\% range across both chips, highlighting the reproducibility with which optimal duty cycle can be achieved with sidewall poling electrodes. 
 
In order to verify the quality of poled domains, a diagonal cross-section of an SPLN waveguide is made by etching a trench through the waveguide after poling. The domains are revealed by immersing the waveguide in an etchant which etches LN crystalline z-faces at different rates (details in Methods) \cite{Xin2024}. The trench is imaged using scanning electron microscopy (SEM) (Fig. \ref{fig:2}d). The images show that the inverted domains extend uniformly across the waveguide ridge as well as all the way into the slab below the ridge. This contrasts with earlier poling techniques that position electrodes only on the slab, restricting domain inversion to the slab region (Fig. \ref{fig:1}e). The experimentally inferred duty cycle throughout the waveguide cross-section reveals little variation (Fig. \ref{fig:2}d plot). This indicates straight and parallel domain walls, ensuring a constant phase matching condition across the entire mode.\\

\subsection*{Experimental results of UV generation}
To accurately estimate the on-chip fundamental and second-harmonic powers we experimentally evaluate the waveguide coupling and propagation losses at 780 and 405 nm, respectively, with a set of spiral waveguides (Fig. \ref{fig:3}a), fabricated using the same process as the SPLN waveguides. The measurements (details in Supplementary Information) show a fiber-to-chip coupling loss of 4.8 dB/facet and a propagation loss of 0.6 dB/cm at 780 nm, achieving a performance comparable to state-of-the-art implementations of visible TFLN waveguides in thinner films than the present work \cite{Renaud2024}. A fiber-to-chip coupling loss of 9.3 dB/facet and a record-low propagation loss of 2.3 dB/cm is measured at 405 nm. 

\begin{figure}[t!]
    \centering
    \makebox[\textwidth][c]{\includegraphics[width=180mm]{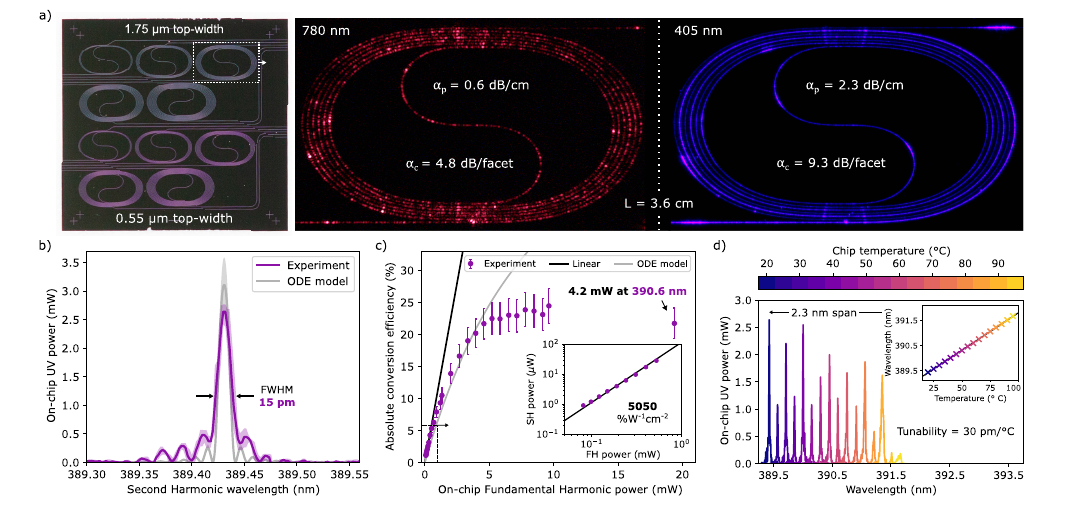}}%
    \caption{\textbf{Characterization of linear and nonlinear properties of fabricated waveguides. }\textit{a) A set of spiral waveguides is used to characterize the propagation ($\alpha_p$) and fiber-to-chip coupling ($\alpha_c$) losses near the fundamental and second harmonic wavelengths, at 780 nm and 405 nm respectively. The record-low propagation loss at 405 nm is attributed to the wide waveguide used, which reduces sidewall scattering loss. b) The phase matching function measured by sweeping the fundamental harmonic wavelength and recording the second harmonic signal. The narrow peak indicates a constant phase matching condition along the length of the 1.5-cm long waveguide and is consistent with the width predicted by our ordinary differential equation (ODE) model. Deviations from the expected $sinc{}^2$-shape function (ODE model) are attributed to residual fabrication errors in the waveguide shape along the waveguide length. The shaded area surrounding the solid lines denotes the deviation in UV signal for both the experiment and theoretical model data. c) Absolute UV conversion efficiency as a function of on-chip power of the fundamental harmonic. At approximately 10 mW of input the conversion efficiency peaks at 24.5\% and a record high on-chip UV power of 4.2 mW is recorded at maximum input power. The grey curve (ODE model) represents a theoretical model that accounts for pump depletion and includes both linear and nonlinear losses (details in main text and Methods). The inset double-logarithmic plot shows the power in the undepleted regime, with a quadratic fit (linear in log-space) indicating a high undepleted conversion efficiency of 5050 \%$\mathrm{W}^{-1}\mathrm{cm}^{-2}$. d) The phase matching function as a function of a chip temperature. A 2.3 nm-wide wavelength tuning range can be achieved.}}
    \label{fig:3}
\end{figure}

Next, we experimentally map the phase matching function (Fig. \ref{fig:3}b), thereby determining the wavelength at which second harmonic generation is maximized. The second harmonic signal as a function of input wavelength is measured by coupling 780 nm light from a mode-hop-free tunable CW laser into the waveguide. At the output of the SPLN chip the transmitted light is filtered to exclude the fundamental harmonic before the second harmonic signal is detected (see Methods). The measured phase matching function features a $\mathrm{sinc{}^2}$-like shape, as predicted by our theoretically derived model, with a narrow ($\sim 15$ pm linewidth) central peak. This indicates that the nonlinear interaction has a near-constant phase matching condition along the entire length of the waveguide, and that the negative impact of fabrication variations is mitigated by adapted poling and our waveguide design. The model is described by a system of coupled mode equations for the second harmonic generation process, including parameters that account for pump depletion and both linear and nonlinear losses, and is discussed with further context below. The slight discrepancy between the measured and theoretically predicted profiles in the sidelobes can be attributed to waveguide geometry imperfections that occur during fabrication and are unaccounted for in the adapted poling process. With the pump wavelength tuned to the peak of the phase matching function the increase of UV power along the propagation length is sufficient for scattered UV light throughout the waveguide to be detected by a camera (Fig. \ref{fig:1}b and c).

In order to characterize the normalized conversion efficiency of the sidewall poled waveguides, the pump wavelength is tuned to the peak of the phase matching function and the second harmonic signal is measured as a function of input power. At sub-mW pump powers, a linear increase in absolute efficiency as a function of pump power is observed (Fig. \ref{fig:3}c). In this undepleted pump regime a record high normalized efficiency of 5050\%W$^{-1}$cm$^{-2}$ is measured (\ref{fig:3}c inset). A theoretically achievable normalized conversion efficiency of 9048\%W$^{-1}$cm$^{-2}$ is calculated using the measured propagation loss at UV and visible wavelengths and nonlinear couplings determined from first principles (details in Methods). This is a factor of $\sim 1.8$ larger than the experimentally measured value, with the discrepancy likely due to a residual phase mismatch. 

Increasing the fundamental harmonic power to around 10 mW a maximum absolute efficiency of 24.5\% is measured. Increasing the power even further we demonstrate a peak on-chip UV power of 4.2 mW at 390.6 nm, an improvement by more than two orders of magnitude over the state of the art \cite{Hwang2023}. We note that this is the first demonstration in TFLN of UV generation at optical powers that exceed the milliwatt level. Accordingly, the leveling of the absolute efficiency in the higher power regime, exceeding pump powers of 3 mW, has not been observed before in this platform. We study this by modeling the interacting field amplitudes using a pair of coupled, ordinary differential equations (ODEs) that describe the second harmonic generation process (see Methods). The model incorporates a fit to the normalized conversion efficiency, measured linear losses, and reported values for UV and visible two-photon absorption (TPA) coefficients in bulk lithium niobate \cite{Beyer2005PRE,Beyer2005OptLett}. Using these parameter values and their reported uncertainties, the model describes the data up to 5 mW of input power (Fig. \ref{fig:3}c), which is well beyond the linear regime. The discrepancy beyond 5 mW is attributed to the fact that the nonlinear loss mechanisms and related effects have not been experimentally characterized in TFLN. A free fit of all TPA coefficients is also explored and matches the experimental data up to higher powers (see Supplementary Information), but gives values which deviate significantly from those previously reported in literature. Therefore, an opportunity is highlighted here for further material investigation in this new wavelength and intensity regime.

\subsection*{Discussion and conclusion}
Using advanced poling strategies for TFLN and an optimized waveguide geometry, we have demonstrated more than 4 mW of UV power generated on-chip, exceeding the state-of-the-art by more than two orders of magnitude. This is enabled by a record high visible-to-UV conversion efficiency of 5050 \%$\mathrm{W^{-1}cm^{-2}}$. Our novel sidewall poling approach allows for complete (100\%) inversion of the TFLN waveguide cross-section using a \textit{pole-after-etch} method that typically can only invert the waveguide slab \cite{Xin2024}. This is particularly effective for waveguide geometries considered in this work that feature very thin slabs (80 nm) and a thick waveguide ridge (520 nm). An early study on different film stacks shows that a 300 nm etched ridge waveguide, on a 600 nm film, can also be completely inverted using sidewall poling albeit at higher poling voltages (details in Supplementary Information). Besides complete domain inversion, we also show that sidewall poling results in straight and parallel domain walls. Precise poling control enabled accurate targeting of the optimal (50\%) poling duty cycle at low poling voltages, verified at multiple locations along the 1.5 cm length waveguide. A saturation of the poling duty cycle - that is, its insensitivity to the precise voltage applied - was observed at higher voltages, which can further simplify the poling process. For example, by pre-compensating the width of the poling fingers a duty cycle around 50\% could be achieved for a wide range of voltages exceeding the saturation voltage. This reduces time consuming poling calibration steps that rely on the iterative process of poling and imaging of the poled devices. 

The high conversion efficiency we demonstrated is also enabled by the record-low loss of 2.4 dB/cm measured near the UV. This propagation loss is comparable to or even lower than that measured in waveguides based on materials with much higher band gaps, such as $\rm Al_2O_3$ and AlN \cite{Franken2023,West2019c,Hendriks2024}. The low propagation loss is attributed to our choice of wide and thick waveguides. While narrower waveguides have an increased mode intensity and thus increased nonlinear interaction \cite{Chen2019}, they suffer from increased phase matching sensitivity (Fig. \ref{fig:2}a) and increased losses from sidewall scattering (see Supplementary Information). For example, we measured the propagation losses of 0.55-$\upmu$m wide waveguides at 405 nm using fabricated spirals (Fig. \ref{fig:3}a) and found a loss value of 5.4 dB/cm. This is much larger than the 2.4 dB/cm measured in the case for the 1.75-$\upmu$m wide waveguides used in this work. As a result, the narrower waveguides have lower theoretical conversion efficiency (8699 \%$\mathrm{W^{-1}cm^{-2}}$) than the wide ones (9048 \%$\mathrm{W^{-1}cm^{-2}}$). This finding was also validated experimentally (Supplementary Information). Further reduction of the sidewall roughness of the waveguides could help decrease the propagation loss and in turn increase the efficiency for narrow waveguides \cite{Chen2023a}.\par

Our poling approach takes into account the film thickness of the starting LN film and etch depth variations \cite{Chen2023,Xin2024}. Using this strategy combined with sidewall poling we obtain a phase matching function close to the ideally expected $\mathrm{sinc{}^2}$ function, indicating high quality poling. Notably, we achieve a 24.5\% absolute conversion efficiency - the highest ever reported for UV generation in this platform - entering a new regime where nonlinear absorption becomes relevant for the first time. The new regime is indicated by a plateau of the conversion efficiency at higher pump powers (Fig. \ref{fig:3}c). We find that our first model using literature values for nonlinear losses (e.g. TPA) matches our experimental data up to about 5 mW. A free fit of model parameters yields a better agreement between experiment and theory (see Supplementary Information), but may result in best-fit TPA coefficients that deviate from literature values. To understand the nonlinear loss mechanisms in TFLN at these wavelengths and intensities, a more in-depth material study is needed. Such studies can also shed light on mechanisms that can help reduce the efficiency diminishing effect of nonlinear losses, such as pulsed pump operation \cite{Beyer2005PRE,Beyer2005OptLett} or optimized waveguide geometries.\par

The mW-level UV generation demonstrated here is sufficient for a wide range of applications from integrated ion traps to microscopy and spectroscopy \cite{Moody2022,Ludlow2015,Lin2022,Wiederkehr2014}. The wide transparency of TFLN, low device complexity and high reproducibility make sidewall poled waveguides in TFLN an excellent candidate for integrated UV sources. While semiconductor laser diodes in the UV require complex process development for shifting the wavelength of interest \cite{Hasan2021,Yang2024}, SPLN waveguides can be adapted for a broad range of wavelengths by simply changing the poling period. Furthermore by heating the SPLN chip, we demonstrate the ability to tune the phase matching function across a 2.3 nm wavelength range, with a tuning efficiency of 30 pm/$\degree$C as expected from literature (Fig. \ref{fig:3}d) \cite{Hwang2023}. The rapid variation in the phase matching function height with respect to wavelength is attributed to mode crossings at elevated temperatures, and the overall trend of decreasing phase matching function height (and increasing width) is attributed to the change in material dispersion with temperature \cite{Schlarb1993}. UV wavelengths beyond the capabilities of our device should be possible straightforwardly using our approach down to 350 nm, the band gap of lithium niobate \cite{Zhu2021a}, albeit with lower efficiency due to increased absorption losses and challenges associated with realizing small poling periods using a lift-off process. Applying sidewall poling to the emerging thin-film lithium tantalate platform, featuring an enhanced photorefraction and optical damage threshold but also reduced nonlinear coefficient, combined with ion-beam milling defined electrodes might become a next step for UV conversion down to 315 nm \cite{Powell2024,Wang2024,Abasahl2021}. Besides wavelength versatility, UV SPLN can potentially bring linewidth reduction to coherent on-chip UV sources where currently intracavity losses limit the laser linewidth \cite{Franken2023, Wunderer2023b}. By upconverting visible on-chip lasers which have widely demonstrated ultra-narrow linewidths \cite{Chauhan2021,Franken2021,Winkler2024,Castro2025}, the laser linewidth is largely transferred to the UV and highly coherent, fully integrated UV sources can be realized \cite{Ling2023}. 

\clearpage
\subsection*{Methods}
\subsubsection*{Nanofabrication and poling}
Alignment markers are first patterned in S1822 resist deposited on a 5\% MgO-doped, x-cut TFLN chip using optical lithography, and then dry-etched into the lithium niobate film. Prior to waveguide patterning, an aligned ellipsometry measurement is performed on a dense grid of points along the intended locations of the waveguides to assess the film thickness non-uniformity. The waveguides, and sets of local alignment markers, are then patterned using aligned e-beam lithography on negative maN-2405 resist and transferred to the lithium niobate film with a reactive Ar-ion etch. A post-fabrication cleanup is performed in heated SC-1 (60 \degree C) to remove etch redeposition, and another aligned ellipsometry measurement is performed to assess the etch depth variations along the chip. To prevent metal contamination of the film and electrically isolate the poling electrodes \cite{Malovichko1999,Nagy2019}, a thin protective $\mathrm{SiO_2}$ interlayer is deposited using inductively coupled plasma chemical vapor deposition.

Using pre-etch and post-etch ellipsometry data, the adapted poling period required for type-0 second harmonic phase matching is calculated and poling electrodes are designed. For the waveguide reported here, the poling period varies from 1.799 $\upmu$m to 1.814 $\upmu$m. To account for lateral spreading of the inverted domains during poling, the width of the poling electrodes are set to 35\% of the poling period. The poling electrodes (Cr/Au) are then fabricated using a lift-off process (LOR 3A and ZEP520A). The electrodes are patterned with an aligned e-beam write using local alignment markers to ensure precise alignment with the pre-fabricated waveguide ridges at a tolerance $\leq$50 nm. After developing, the electrodes are defined using e-beam evaporation and lift-off. First, an adhesion layer of chromium (25 nm) is deposited, followed by a layer of gold (125 nm). Following this step, the remaining resist is lifted off to define the electrodes. A layer of photoresist (S1822) is deposited on top of the electrodes to prevent breakdown during the poling, and contact windows for the poling probes are defined using photolithography. Poling is performed by touching probes onto the exposed pads and by applying a voltage pulse. The poling pulse parameters can be found in \cite{Xin2024}. The pulses are generated by a digital-to-analog converter and then amplified to reach the required voltage. Non-destructive imaging using oil-immersed second harmonic microscopy is used to verify the poling at a high resolution, for at least 3 locations along each waveguide.

After poling, the chip facets are defined using a deep etching process. First, the protective photoresist is removed by immersing the sample into Remover PG, the electrodes are etched away in gold etchant followed by chromium etchant, and finally the oxide interlayer is removed using diluted hydrofluoric acid. Then, 8.5-$\upmu$m thick photoresist (SPR-220-7.0) is deposited, and boundaries for the deep etch are defined using photolithography. After developing, the lithium niobate and buried oxide are fully etched through at the facet locations using a reactive ion etch, and the silicon handle is etched through using a Bosch process etch \cite{He2019}. Finally, any remaining resist is removed from the singulated chip by immersing in Remover PG. The chip is finally annealed for 2 hours at 520$^\circ$C in an oxygen atmosphere to reduce absorption losses.

To inspect the quality of our poling, we produce the structure as in Fig. \ref{fig:2}d using the following process. First, a thin $\mathrm{SiO_2}$ layer is deposited on the SPLN waveguide using plasma enhanced chemical vapor deposition. Next, a diagonal pattern is defined using photolithography at an angle that should cross the waveguide width in about 20 poling periods and etched using reactive ion etching. To expose the inverted and non-inverted domains in the SPLN waveguide, revealed by the diagonal etch, the sample is immersed in heated SC-1 at 65 $\degree$C, taking advantage of the fact that SC-1 etches the faces of inverted and non-inverted regions of the crystal with different rates. Next, the protective $\mathrm{SiO_2}$ layer is removed in diluted hydrofluoric acid, and the sample is imaged using scanning electron microscopy.

 \subsubsection*{Experimental set-up for UV SPLN characterization}
The SPLN chip is mounted on a copper stage that is temperature controlled using a Peltier element and a thermistor. During the measurement of the phase matching function (Fig. \ref{fig:3}b) the chip was at a temperature of 20 \degree C. For all data points in Fig. \ref{fig:3}c the chip was set to 65 \degree C. Fiber-coupling for the input and output facet is achieved with lensed PM-630HP fibers (OZ Optics) and two high precision, three-axis stages (Thorlabs NanoMax 300). We do not find any difference measuring power or phase matching function curves using this fiber (effectively multimode for UV) and a single mode UV fiber (Thorlabs SM300). This indicates our UV generated light is dominated by the single $\mathrm{TE_{00}}$ mode. We chose to perform measurements with PM-630HP instead of a single mode UV fiber because the former also guides the pump wavelength, and thus can facilitate setup alignment even in the absence of UV light. No photodarkening or solarization of this fiber was observed during the experiments. The pump power exiting the lensed fiber and incident on the chip is measured with a calibrated visible range photodiode (Thorlabs S120C, with fiber mount) and the UV power at the output fiber is collected using a calibrated UV photodiode (Thorlabs S120VC, with fiber mount); for the latter the pump is blocked with a filter (OD $\mathrm{>}$7 at 780 nm, 405/150 nm BrightLine single-band bandpass filter). For measurements of the phase matching function a variable gain amplified photodiode (Thorlabs PDA100A2, with fiber mount) is used in combination with the previously mentioned pump filter to measure the UV power. The phase matching function is recorded by sweeping the laser (CW, Toptica DL Pro 780 nm), while recording the laser wavelength and power (HighFinesse Wavelength Meter WS-7), and the UV signal on the photodiode using a digital-to-analog converter (National Instruments USB-6210). The pump laser can be swept mode-hop-free reliably over about 85 GHz frequency range, using both piezo and current control of the laser. The recording for a typical phase matching function measurement spans about 400 GHz and is constructed by stitching mode-hop-free sweeps together while rotating the grating of the laser simultaneously (with an automated motor control). The data analysis method to obtain the phase matching function from the raw measurement data can be found in the Supplementary Information. For the measurements of the absolute conversion efficiency as a function of pump power (Fig. \ref{fig:3}c) a variable optical attenuator is used (Thorlabs V600PA), except for the highest pump power measurement. The insertion loss of the attenuator is about 3 dB, explaining the gap between the highest and second highest fundamental harmonic power data point in Fig. \ref{fig:3}c.

\subsubsection*{Sensitivity calculation}
For a given periodic poling grating momentum $G$ and waveguide top width $w$ the phase matched second harmonic generation wavelength is whichever second harmonic wavelength $\lambda_{\mathrm{opt}}$ minimizes the phase mismatch $\Delta \beta_\mathrm{QPM} = 2\beta(\lambda_\mathrm{FH},w) - \beta(\lambda_\mathrm{SH},w)$:

\begin{equation} \label{eq:lambda_opt}
    \lambda_\mathrm{opt}(w) = \argmin \left( \left| 2\beta(\lambda_\mathrm{FH},w) - \beta(\lambda_\mathrm{SH},w) + G \right| \right)
\end{equation} 

where $\lambda_\mathrm{SH}$ is the second harmonic wavelength and $\lambda_\mathrm{FH} = 2\lambda_\mathrm{SH}$ is the fundamental harmonic wavelength. Ideally $\lambda_\mathrm{opt}$ would be the intended design wavelength, but due to variation in waveguide top width and the resulting change in geometric dispersion, the optimally phase matched SH wavelength often differs from that intended. For a fixed geometry and a given $G$ and nominal film thickness $w = w_0$, then, the phase matching sensitivity is the quantity $\frac{d\lambda_{\mathrm{opt}}}{dw} \Bigr|_{w_0}$. We approximate this derivative by simulating the interacting modes in the nominal and slightly perturbed geometries. A more detailed description and derivation can be found in the Supplementary Information.

\subsubsection*{Theoretical estimation of the conversion efficiency}
To obtain a semi-analytical expression for the conversion efficiency in our second-harmonic generation experiments, we assume that the fundamental harmonic and second harmonic field experience only linear losses and no back-conversion from the generated UV back to the pump takes place. Under these approximations, the conversion efficiency ($\eta_{\text{abs}}$) can be written as

\begin{equation}
\label{eq:theoryefficiency}
    \eta_{\text{abs}} := \left|\frac{a_{2}(L)}{a_{1}(0)}\right|^{2} = \frac{4}{\pi^{2}}\left|\kappa_{\text{eff}}\right|^{2}\left|a_{1}(0)\right|^{2}L^{2}e^{-\left(\alpha_{1}+\frac{\alpha_{2}}{2}\right)L}\left(\frac{\sinh^{2}\left[\left(\alpha_{1} - \frac{\alpha_{2}}{2}\right)\frac{L}{2}\right]}{\left[\left(\alpha_{1} - \frac{\alpha_{2}}{2}\right)\frac{L}{2}\right]^{2}}\right).
\end{equation}

Here, \(a_{1}(0)\) is the amplitude of the fundamental harmonic field, \(a_{2}(L)\) is the amplitude of the second harmonic (UV) field at the waveguide output \(z = L\), \(\kappa_{\mathrm{eff}}\) is the effective coupling constant proportional to the second-order susceptibility \(d_{zzz}\), \(\alpha_{1}\) and \(\alpha_{2}\) are the measured linear propagation losses for the fundamental and second harmonic fields, respectively, and \(L\) is the length of the poled waveguide. The exponential factor accounts for amplitude attenuation due to losses, and the hyperbolic sine term captures the interplay between pump and second-harmonic attenuation rates. More details can be found in the Supplementary Information.

\subsubsection*{Ordinary differential equation (ODE) model}
The coupled differential equations used to model frequency conversion from the fundamental field $a_1$ to the second harmonic field $a_2$ as a function of propagation length $z$ incorporating linear and nonlinear losses are

\begin{equation}
    \frac{da_1}{dz} = -i\kappa a_2 a_1^* e^{i\Delta\beta z}-\frac{\alpha_{1}}{2}a_1-\frac{1}{2}\left(\frac{\beta_{11}}{A_\mathrm{eff,1}}\left|a_1\right|^2+\frac{\beta_{12}}{A_\mathrm{eff,mix}}\left|a_2\right|^2\right)a_1
\end{equation}
\begin{equation}
    \frac{da_2}{dz} = -i\kappa \left|a_1\right|^2 e^{-i\Delta\beta z}-\frac{\alpha_{2}}{2}a_2-\frac{1}{2}\left(\frac{\beta_{21}}{A_\mathrm{eff,mix}}\left|a_1\right|^2+\frac{\beta_{22}}{A_\mathrm{eff,2}}\left|a_2\right|^2\right)a_2
\end{equation} 

where $\kappa$ is the second harmonic generation coupling, $\Delta \beta$ is the phase mismatch, $\alpha_{i}$ are the measured linear absorption coefficients, $\beta_{ij}$ are the two photon absorption coefficients for photons from $a_i$ and $a_j$, $A_\mathrm{eff,j}$ are the effective mode areas of the fields, and $A_\mathrm{eff,mix}$ is the average of the effective mode areas of $a_1$ and $a_2$.

These equations are solved numerically to determine $a_2$ as a function of input fundamental power. The parameters are allowed to vary within experimental error in order to fit the experimental data.  More details can be found in the Supplementary Information.

\clearpage
\subsection*{Acknowledgments}
We thank N. Sinclair for providing the fiber infrastructure for our experimental set-up. We thank Y. Song and A. Shelton for loaning experimental equipment. We were grateful for helpful discussions on fabrication with L. Magalhaes, K. Powell, and X. Zhu. We thank A. McClelland for technical support and guidance on SH microscope imaging. Devices were fabricated in the Center for Nanoscale Systems (CNS) at Harvard.

\subsection*{Contributions}
C.A.A.F, S.S.G, C.C.R. and C.J.X. conceived the idea of sidewall poling. C.A.A.F, S.S.G, C.C.R. and J.Y. contributed to the theoretical modeling and design of the waveguides and sidewall electrodes. C.A.A.F, S.S.G, C.C.R., J.Y., C.J.X., S.L. and D.W. contributed to the fabrication of the UV SPLN devices. C.A.A.F and D.W. contributed to SEM and camera imaging of the fabricated structures. C.A.A.F., S.S.G. and J.Y. contributed to poling and second harmonic imaging of the devices. C.A.A.F., S.S.G. and G.J. contributed to experimental set-up design. C.A.A.F. performed  measurements of UV generation. C.A.A.F. and S.S.G. drafted the manuscript, with G.S.W., K.J.B., and M.L. contributing to the writing. Comments and feedback from all co-authors were incorporated.

\subsection*{Funding}
Dutch Research Council (NWO) under the grant "Ultra-narrowband lasers on a chip" (16718); Department of the Air Force (DAF) DoD FA9453-23-C-A039; Amazon Web Services (A50791); Office of Naval Research (N00014-22-C-1041/VAS-21-0001); NASA (80NSSC22K0262, 80NSSC23PB442 ); NSF (ERC EEC-1941583,  OMA-2137723); National Research Foundation funded by the Korea government (NRF-2022M3K4A1094782). S.L. acknowledges  A*Star – fellowship;

\subsection*{Disclosures}
M.L. is involved in developing lithium niobate technologies at HyperLight Corporation.

\subsection*{Data availability}
The data that support the findings of this study are available from the corresponding author upon reasonable request.

\subsection*{Code availability}
The code that support the findings of this study are available from the corresponding author upon reasonable request.
\clearpage

\newpage
\supplementarysection
\clearpage

\subsection*{Phase matching sensitivity calculation}
The phase matching sensitivity is defined as the derivative of the optimally phase matched second harmonic (SH) wavelength (in nm) with respect to waveguide top width (also measured in nm). For a given SH wavelength $\lambda_{\mathrm{SH}}$, the second harmonic generation phase mismatch is given as $\Delta \beta(\lambda_\mathrm{SH},w,...) = 2\beta(\lambda_\mathrm{FH},w,...) - \beta(\lambda_\mathrm{SH},w,...)$, where $\lambda_\mathrm{FH} = 2\lambda_\mathrm{SH}$ and the mode propagation constants $\beta$ are determined  by both the material dispersion of the crystal and the geometric dispersion of the waveguide. The geometry of the waveguide is mainly parametrized by the waveguide top width $w$, film thickness, etch depth and sidewall angle. In this treatment we focus on the effect of the top width $w$ specifically. The presence of a constant poling period $\Lambda$ contributes a fixed quasi-phase matched grating momentum $G = 2\pi/\Lambda$, which we then write overall as $\Delta \beta_\mathrm{QPM}(\lambda_\mathrm{SH},w) = 2\beta(\lambda_\mathrm{FH},w) - \beta(\lambda_\mathrm{SH},w) + G$. For a given $G$ and $w$ the phase matched SHG  wavelength ($\lambda_{\mathrm{opt}}$) is the one that  minimizes $\Delta \beta_\mathrm{QPM}$:

\begin{equation} \label{eq:lambda_opt}
    \lambda_\mathrm{opt}(w) = \argmin \left( \left| 2\beta(\lambda_\mathrm{FH},w) - \beta(\lambda_\mathrm{SH},w) + G \right| \right)
\end{equation}

However, due to waveguide width nonuniformity,  which  impacts the geometric dispersion,  the optimally phase matched second harmonic wavelength often differs from this value. For an otherwise fixed geometry and a given $G$ and nominal waveguide width $w = w_0$, then, the phase matching sensitivity is defined as  $\frac{d\lambda_{\mathrm{opt}}}{dw} \Bigr|_{w_0}$. To evaluate this sensitivity  numerical modeling is used to calculate $\lambda_{\mathrm{opt}}$ for different waveguide top widths,  assuming particular values for film thickness, etch depth, and quasi-phase matched grating momentum $G$. Even though film thickness and etch depth variations can contribute to sensitivity, they are not considered here since these variations are accounted for by the adapted poling method used. However, an earlier study \cite{Hwang2023}, confirmed by our own calculations in the text below, concluded that thick and wide waveguides have the lowest sensitivity to film thickness variations. 

The derivative is calculated numerically. Specifically, we calculate $(w_{-},\lambda_\mathrm{opt}^{-}), (w_{0},\lambda_\mathrm{opt}^0)$, and $(w_{+},\lambda_\mathrm{opt}^{+})$, where $w_{\mp} = w_0 \mp \delta w$ are small perturbations from the nominal waveguide width $w_0$ and $\lambda^{\mp}_\mathrm{opt}$ are the corresponding values for phase matched SHG wavelength. The slope of the least mean squares linear fit between these three points is then taken as an approximation to $\frac{d\lambda_{\mathrm{opt}}}{dw} \Bigr|_{w_0}$. 

\begin{figure}[h!]
    \centering
    \makebox[\textwidth][c]{\includegraphics[width=60mm]{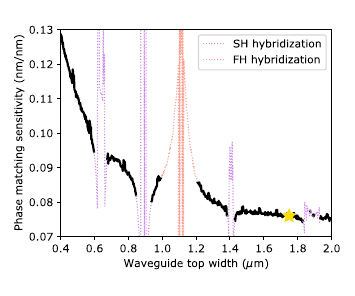}}
    \caption{\textit{Phase matching sensitivity with respect to film thickness variation, for different waveguide top widths. The star denotes the chosen waveguide width of 1.75 $\mu$m}}
    \label{fig:sensitivitythickness}
\end{figure}

Similarly, the phase matching sensitivity is also evaluated for variations in film thickness. For a given film thickness $t = t_0$ the sensitivity is defined as $\frac{d\lambda_\mathrm{opt}}{dt}\Bigr|_{t_0}$ and evaluated for a range of waveguide top widths (Fig. \ref{fig:sensitivitythickness}). Like in Fig. \ref{fig:2}a in the main text, we find that wider waveguides have a lower phase matching sensitivity to fabrication tolerances (in this case to film thickness variations).

\clearpage

\subsection*{Waveguide propagation loss and fiber-to-chip coupling loss}
The propagation loss and fiber-to-chip coupling loss for the fiber-to-chip coupling is measured by fabricating a set of spiral waveguides of varying length and measuring the transmission at the wavelengths of interest. The set-up is described in the Methods of the main text. The fibers used here are PM-630HP fibers (OZ Optics). For transmission measurements at 780 nm a CW Toptica DL Pro 780 laser is used. For transmission measurements at 405 nm a CW, fiber-pigtailed, Fabry-Pérot laser diode is used (QPhotonics QFLD-405-30SAX-PM) in combination with a laser diode driver (Thorlabs CLD1010LP). The power is measured using a calibrated photodiode (Thorlabs S150C). To interpret our measurements we assume a symmetric fiber-to-chip coupling loss. We support this assumption by verifying that the incident beam from the lensed fiber is single transverse mode for both wavelengths.\\

\begin{figure}[h!]
    \centering
    \makebox[\textwidth][c]{\includegraphics[width=180mm]{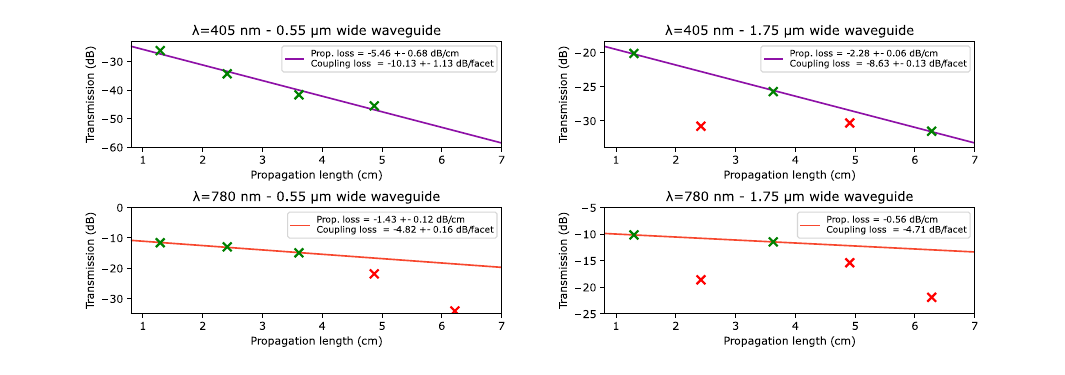}}
    \caption{\textit{Transmission measurements for waveguide spirals of various lengths with two waveguide top widths, 0.55 and 1.75 $\mu$m, measured at two wavelengths, 405 and 780 nm. The red markers indicate spiral measurements where the transmission measurement is not accurate due to a fabrication error in the spiral structure (see also Fig. \ref{fig:Spiral405nm_175wide} to \ref{fig:Spiral780nm_055wide}). Only the green marked measurements are used for the linear fit and loss extraction.}}
    \label{fig:propfacetloss}
\end{figure}

\begin{figure}[h!]
    \centering
    \makebox[\textwidth][c]{\includegraphics[width=180mm]{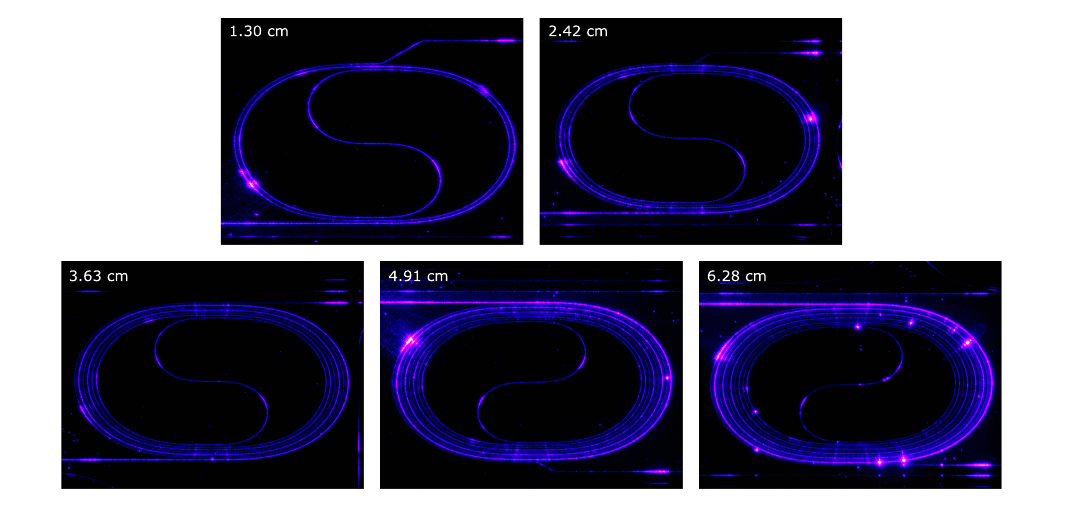}}%
    \caption{\textit{Top-down microscope images of 1.75 $\mu$m wide waveguide spirals with 405 nm light at the input.}}
    \label{fig:Spiral405nm_175wide}
\end{figure}

\begin{figure}[h!]
    \centering
    \makebox[\textwidth][c]{\includegraphics[width=180mm]{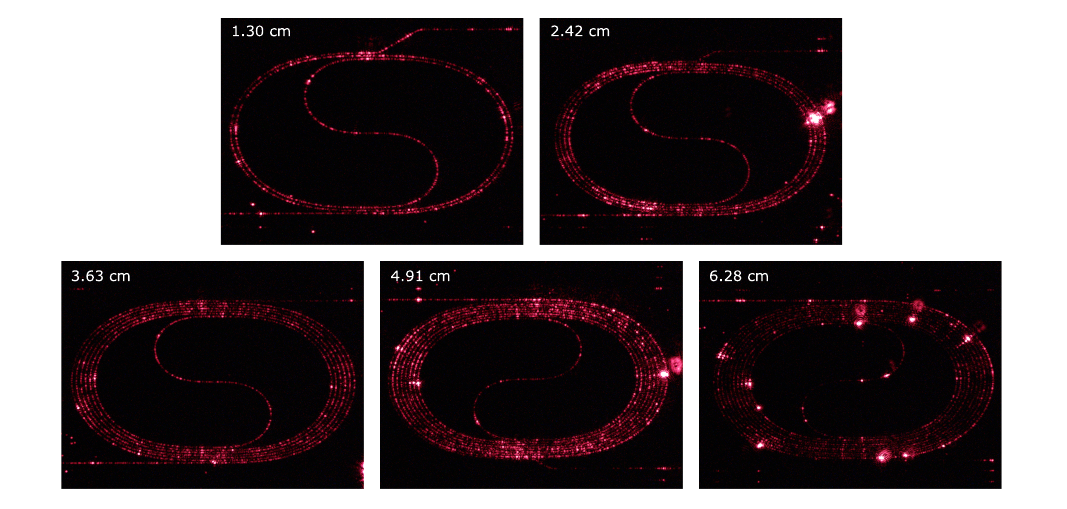}}
    \caption{\textit{Top-down microscope images of 1.75 $\mu$m wide waveguide spirals with 780 nm light at the input.}}
    \label{fig:Spiral780nm_175wide}
\end{figure}

\begin{figure}[h!]
    \centering
    \makebox[\textwidth][c]{\includegraphics[width=180mm]{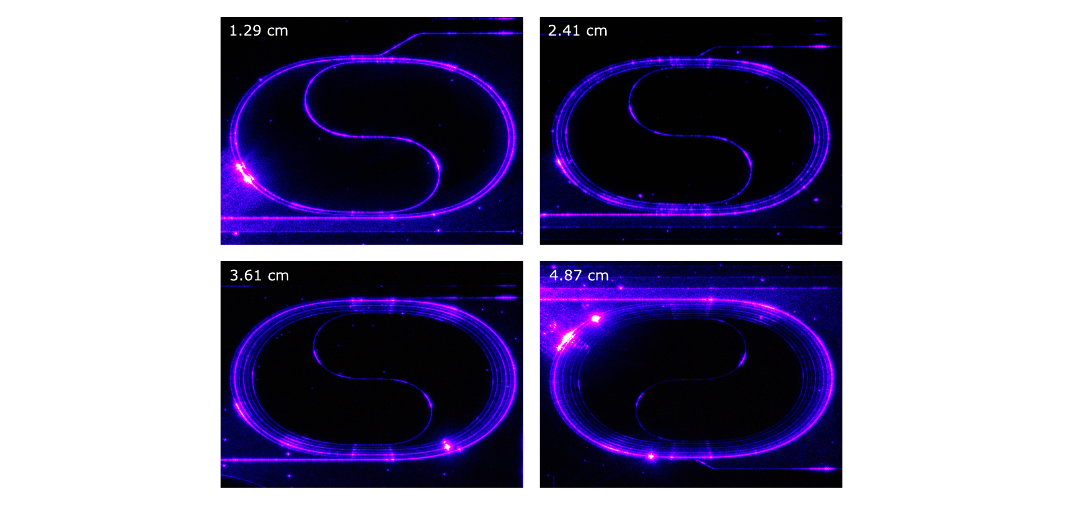}}
    \caption{\textit{Top-down microscope images of 0.55 $\mu$m wide waveguide spirals with 405 nm light at the input. The fifth spiral of 6.22 cm did not transmit any measurable 405 nm light, no image was taken for this spiral.}}
    \label{fig:Spiral405nm_055wide}
\end{figure}

\begin{figure}[h!]
    \centering
    \makebox[\textwidth][c]{\includegraphics[width=180mm]{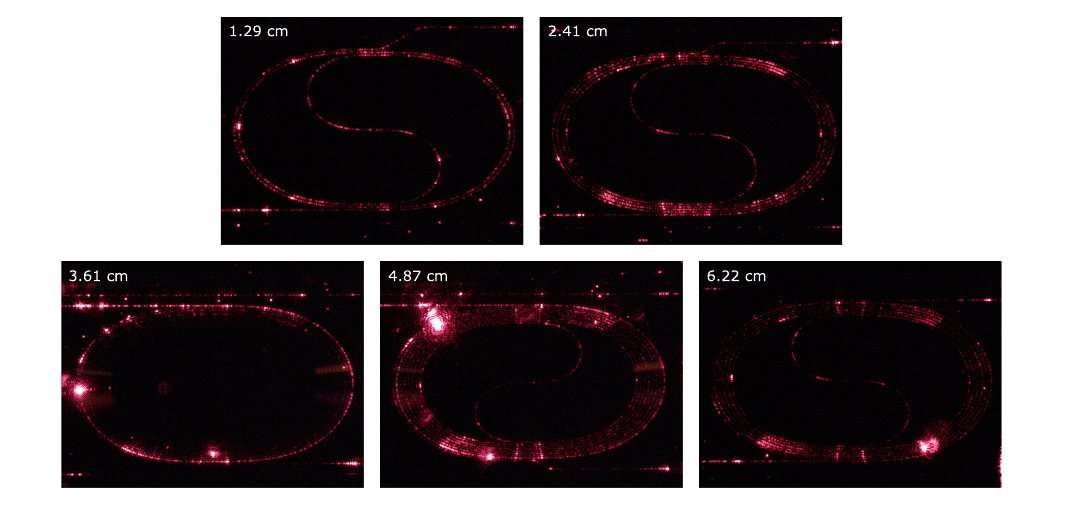}}
    \caption{\textit{Top-down microscope images of 0.55 $\mu$m wide waveguide spirals with 780 nm light at the input.}}
    \label{fig:Spiral780nm_055wide}
\end{figure}

The transmission as a function of propagation length through the spiral structures, for both wavelengths, is shown in Fig. \ref{fig:propfacetloss}. Measurements of spirals with defects, as can also be seen as bright scatter points in the images in Fig. \ref{fig:Spiral405nm_175wide} to \ref{fig:Spiral780nm_055wide}, are excluded from the linear fit that estimates the propagation and fiber-to-chip coupling loss. As the facets of both chips are the same, the average coupling loss can be determined from spirals of both waveguide widths. We find an average fiber-to-chip coupling loss of 9.32 $\mathrm{\pm}$ 0.48 dB and 4.77 $\mathrm{\pm}$ 0.37 dB for 405 nm and 780 nm, respectively.

\FloatBarrier

\newpage
\subsection*{Film thickness and etch depth variation}
\begin{figure}[h!]
    \centering
    \makebox[\textwidth][c]{\includegraphics[width=120mm]{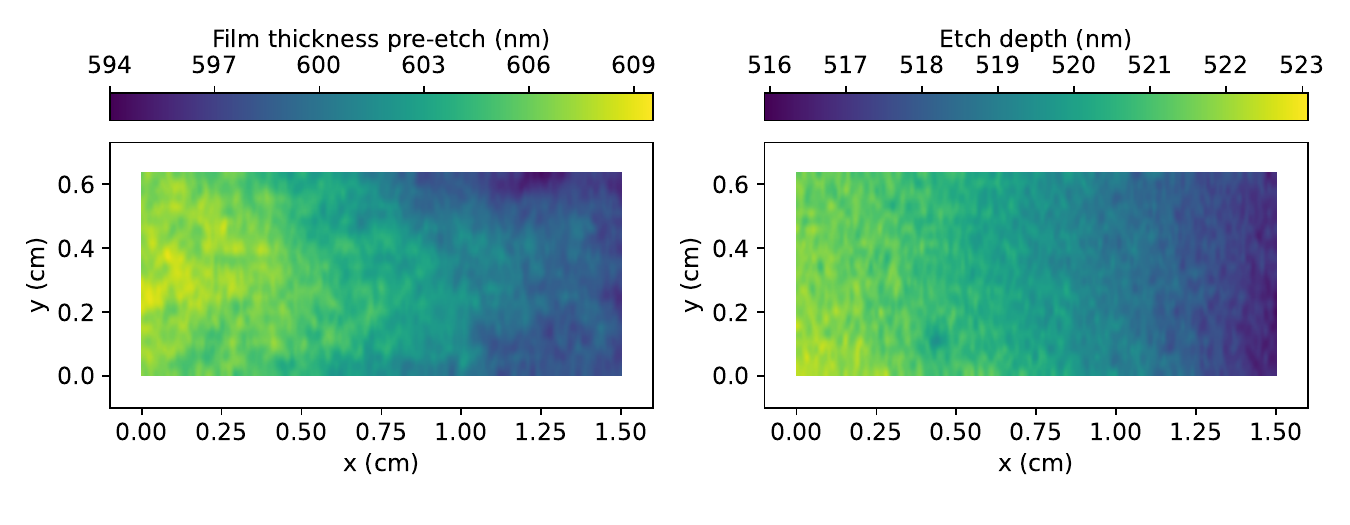}}
    \caption{\textit{Left, lithium niobate film thickness before etching the waveguides. Right, etch depth measured after waveguide definition. Here, the etch depth is the film thickness after etching subtracted from the film thickness before etching.}}
    \label{fig:filmthickness}
\end{figure}

In our adapted poling process the film thickness and etch depth are taken into account to calculate the optimal poling period along the waveguide. The film is measured, before and after etching, using a Woollam RC2 Spectroscopic Ellipsometer. The horizontal spacing between data points is 100 $\upmu$m, measured at the center of each waveguide. The results are shown in Fig. \ref{fig:filmthickness}.

\clearpage

\newpage
\subsection*{Optical  intensity at the waveguide sidewall}
\begin{figure}[h!]
    \centering
    \makebox[\textwidth][c]{\includegraphics[width=115mm]{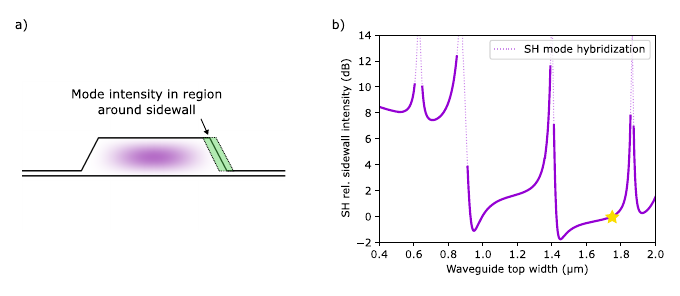}}
    \caption{\textit{a) During the simulation the mode intensity is calculated in an area surrounding the waveguide sidewall. b) Relative sidewall intensity of the second harmonic UV mode as a function of waveguide width. The intensity is related to the intensity at our waveguide width of 1.75 $\mu$m, indicated by the star.}}
    \label{fig:sidewall}
\end{figure}

Several factors contribute to the total waveguide propagation loss, such as linear absorption, nonlinear (two-photon) absorption, and sidewall scattering. The last factor is highly dependent on the waveguide geometry and mode confinement, since the intensity of the optical mode on the sidewall is directly proportional to the scattering loss. Scattering loss from sidewall roughness is proportional to $1/\lambda^3$ \cite{Melati_2014jo}, therefore we will only consider the dominant scattering loss from the second harmonic UV mode here. To investigate what waveguide geometry is optimal, the sidewall intensity for the second harmonic (SH) mode is calculated for various waveguide widths using COMSOL (Fig. \ref{fig:sidewall}a and b). At some waveguide widths the fundamental mode has a TE polarization below 90\%, these regions are labeled as regions with hybridized modes and indicated with a dotted line. The trend of the solid line shows that wider waveguides have lower sidewall intensity, thus lower sidewall scattering loss.

\newpage

\subsection*{Data analysis to obtain the duty cycle from SH microscope images}
\begin{figure}[h!]
    \centering
    \makebox[\textwidth][c]{\includegraphics[width=180mm]{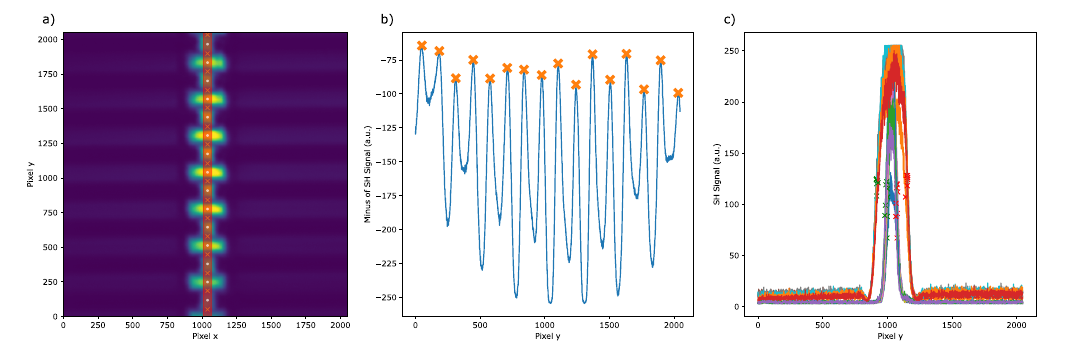}}
    \caption{\textit{a) Second harmonic microscope image, here the SH signal, of a sidewall poled lithium niobate waveguide with the poling fingers still on the waveguide. The red shaded area denotes what part of the image will be used to determine the duty cycle. b) The data in the red shaded area (see a) averaged along x multiplied by -1. Peaks in this line (orange cross-markers) correspond to dark points/lines in the SH microscope image, which indicate a boundary of the poled domains in lithium niobate. c) Cross-sections of the image (along x) at the white dots in a). This allows us to determine whether a domain is inverted or not inverted by comparing the widths of the curves. The inverted domains are partially obscured by the poling fingers which reduces their width in these cross-sectional plots. Using this we can determine which part of the image is inverted and whether the duty cycle is below or above 50\%. The duty cycle reported is the mean duty cycle extracted and the error is the standard deviation of the duty cycle extracted from the image through the peak finding algorithm. The extracted duty cycle for this image is 51.39 $\pm$ 0.96 \% for a poling voltage of 139 V.}}
    \label{fig:SHmicroscope}
\end{figure}

A high-resolution second harmonic microscope with an oil-immersion objective is used to determine the duty cycle of our poled structures, as described in the main text. For chip \#1 and \#2 we have recorded 25 and 26 second harmonic microscope images, respectively. A single data point sacrifices a single test structure, since these can be poled reliably only once. For the results shown in the main text (Fig. \ref{fig:2}c) a total of 51 test structures were used, each about 500 $\upmu$m in length with a poling period that is comparable to what is used for the 1.5 cm long, functional SPLN waveguides. The analysis to obtain the duty cycle from a single device is described in the caption of Fig. \ref{fig:SHmicroscope}.

\clearpage

\newpage
\subsection*{Calibration of the phase matching function}
\begin{figure}[h!]
    \centering
    \makebox[\textwidth][c]{\includegraphics[width=180mm]{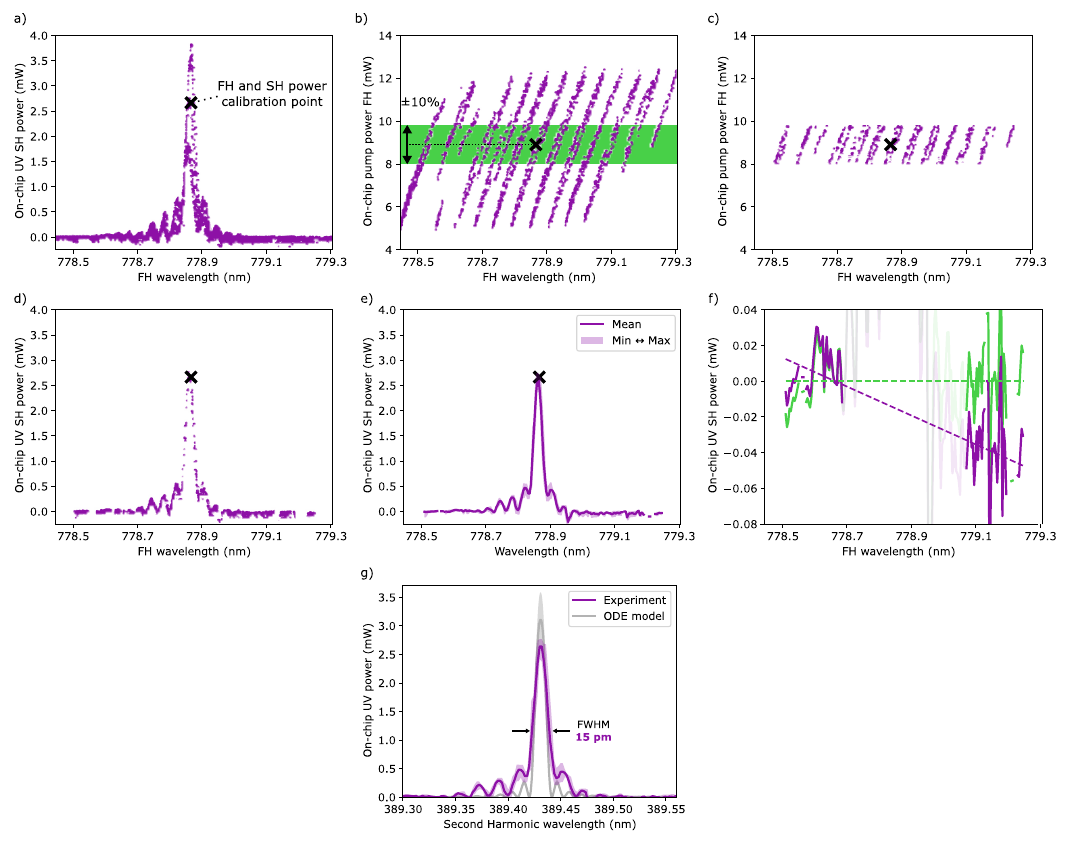}}
    \caption{\textit{Data analysis to obtain the phase matching function from the raw experimental data. See text in this section for a subfigure description.}}
    \label{fig:explanationPMF}
\end{figure}

The experimental setup and measurement procedure used to obtain the phase matching function are described in detail in the Methods section of the main text. Here, we focus on how the raw photodiode data is processed to produce the phase matching curve shown in Fig. \ref{fig:3}b of the main text. Each step of this procedure is illustrated in Fig.~\ref{fig:explanationPMF}:

\begin{enumerate}[label=\textbf{(\alph*)}]
\item \textbf{Raw photodiode trace and calibration.}  
As the fundamental harmonic (FH) pump laser is swept in wavelength across the phase matching region, the output of the UV SPLN waveguide is measured on a UV photodiode (with a pump filter). The photodiode voltage is recorded as a function of the FH wavelength. After the sweep, at the peak-conversion wavelength, the UV power is measured with a calibrated power meter. This single calibration point---mapping photodiode voltage (V) to optical power (mW)---is then applied to convert the entire photodiode trace into UV power. In the example shown, the on-chip UV power at peak efficiency is 2.7\,mW (corresponding to 1.60\,V on the photodiode), indicated by the black cross in subfigure~(a). This occurs at an on-chip pump power of 8.90\,mW and an FH wavelength of 778.87\,nm.

\item \textbf{Pump power–based data selection.}
A 99/1\,\% fiber splitter is placed at the pump laser output, sending 1\,\% of the FH light to a wavelength meter that also measures input power. Because the laser power fluctuates during the sweep, only data points whose pump power lies within $\pm10\,\%$ of the reference power measured at the calibration point (the black cross) are retained.

\item \textbf{Selected data pump power vs.\ wavelength.}
After applying this $\pm10\,\%$ pump power selection, the remaining on-chip FH pump power is plotted as a function of the FH wavelength.

\item \textbf{Selected data UV power vs.\ wavelength.}
The corresponding UV powers, for the same selected data points, are plotted versus FH wavelength.

\item \textbf{Binning and averaging.}
To reduce noise from alignment drifts (likely caused by small fiber vibrations), the data are grouped into 4\,pm wavelength bins and averaged (the solid line). The shaded region indicates the minimum-to-maximum spread within each bin. Some bins contain no data points (due to the $\pm10\,\%$ filtering) but lie outside the main phase-matched region (e.g., at 779.2\,nm), so they do not affect the overall shape of the phase matching function.

\item \textbf{Offset drift correction.}
Over the 3--4\,min duration of the measurement, a small DC-voltage drift in the photodiode (on the order of 10\,mV) could lead to negative power readings. To correct for this, the photodiode noise floor is measured at the start and end of the sweep, and its slope is subtracted from the dataset so that the noise floor remains near zero throughout.

\item \textbf{Final comparison with theory.}
The fully processed data are plotted alongside results from the coupled ordinary differential equation (ODE) model. The measured on-chip pump power and wavelength serve as inputs to the model; the band around the theoretical curve shows the effect of the $\pm10\,\%$ pump-power variation on the UV output power.
\end{enumerate}

\clearpage

\subsection*{Theoretical estimation of the normalized conversion efficiency}
We model second harmonic generation (SHG) using the following dynamic equations that describe the  propagation of the waves along the waveguide oriented  in the $z$ direction (aligned with the $Y$ material axis).

\begin{equation}\label{eq:dynamic_equations_red_UV}
    \frac{da_{1}}{dz} = -i\kappa^{*}a_{1}^{*}a_{2}e^{i \Delta \beta z} - \frac{\alpha_{1}}{2}a_{1}  \quad \text{and} \quad \frac{da_{2}}{dz} = -i\kappa a_{1}^{2}e^{-i \Delta \beta z}  - \frac{\alpha_{2}}{2}a_{2}.
\end{equation}

The functions $a_{1}(z)$ and $a_{2}(z)$ are the electric field intensities for the fundamental and second harmonic fields, respectively. The parameter $\Delta \beta  = \beta_{2} - 2 \beta_{1}$ is the phase mismatch between the wavevectors $\beta _{1}$ and $\beta _{2}$; $\alpha_{1}$ and $\alpha_{2}$ are the linear propagation losses; and $\kappa$ is the coupling constant for the SHG interaction. The equations are symmetrically coupled when normalized to intensity due to the degeneracy of the upconversion process $2\omega = \omega + \omega$ and the non-degeneracy of the simultaneous downconversion process $\omega = 2\omega - \omega$. This system does not have an analytic solution. Therefore, to obtain a semi-analytical expression to guide our intuition for the conversion efficiency, we assume that the pump intensity ($a_{1}$) is independent of the SHG interaction. That is, the pump only experiences linear losses, and there is no back-conversion from the second harmonic to the fundamental harmonic. Thus

\begin{equation}
    \frac{da_{1}}{dz} \approx - \frac{\alpha_{1}}{2}a_{1} \quad \Rightarrow \quad a_{1}(z) = a_{1}(0)e^{-\frac{\alpha_{1}}{2}z}
\end{equation}

the evolution of the second harmonic is then given by'

\begin{equation}
    \frac{da_{2}}{dz} \approx - \frac{\alpha_{2}}{2}a_{2} -i\kappa \left(a_{1}(0)e^{-\frac{\alpha_{1}}{2}z}\right)^{2}e^{-i\Delta \beta z}.
\end{equation}

The idea of periodic poling is precisely to correct the phase mismatch represented by the exponential $e^{-i\Delta \beta z}$ in Eq. \ref{eq:dynamic_equations_red_UV}. To eliminate this term, we modulate $\kappa$ in such a way that $\kappa \sim e^{i\Delta \beta z}$.  $\kappa$ is proportional to the second-order nonlinear susceptibility $d_{ijk}$ through the equation

\begin{equation}
    \kappa = \frac{\varepsilon_{0}\omega_{\text{SH}}}{v_{g,\text{SH}}}\left(\int_{S}\varepsilon_{ij}E_{2}^{i}E_{2}^{j}dS\right)^{-1}\int_{S}d_{ijk}E_{2}^{i}E_{1}^{j}E_{1}^{k}dS,
\end{equation}

where $E_{1}^{i}$ is the $i$-th component of the fundamental harmonic; $E_{2}^{i}$ is the $i$-th component of the second harmonic; $\varepsilon_{0}$ is the vacuum electric permittivity; $\varepsilon_{ij}$ is the $(i,j)$-th component of the electric permittivity tensor; $v_{g,\text{SH}}$ is the group velocity of the second harmonic field and $\omega_{\text{SH}}$ is the angular frequency of the second harmonic field. We can now directly modulate $d_{iJ}$ (using contracted Voigt notation for the indices) in order to achieve phase matching as follows

\begin{equation}
    d_{iJ}(z) \approx
    \begin{pmatrix}
        0 & 0 & 0 & 0 & 0 & 0 \\
        0 & 0 & 0 & 0 & 0 & 0 \\
        0 & 0 & d_{33} & 0 & 0 & 0
    \end{pmatrix}
    \text{sign}\left[\cos{\left(\Delta \beta z\right)}\right]
    =
    d_{zzz}\text{sign}\left[\cos{\left(\Delta \beta z\right)}\right]\delta_{iz}\delta_{jz}\delta_{kz}
\end{equation}
where the $\delta_{iz}, \delta_{jz}, \delta_{kz}$ are Kronecker delta functions. Here,  since most of the SHG comes from the $d_{33} = d_{zzz}$ coefficient, we  assumed  that only this coefficient is non-zero.  The choice of a square wave modulation is supported by our experimental data that indicate that our domains are square-shaped. Then, the coupling parameter $\kappa$ becomes

\begin{equation}
    \kappa = \underbrace{\left(\frac{\varepsilon_{0}\omega_{\text{SH}}}{v_{g,\text{SH}}}\frac{\int_{S}d_{zzz}E_{2}^{z}E_{1}^{z}E_{1}^{z}dS}{\int_{S}\varepsilon_{ij}E_{2}^{i}E_{2}^{j}dS}\right)}_{:=\hspace{0.05cm}\kappa_{\text{eff}}}\text{sign}\left[\cos{\left(\Delta \beta z\right)}\right] := \kappa_{\text{eff}}\hspace{0.05cm}\text{sign}\left[\cos{\left(\Delta \beta z\right)}\right],
\end{equation}

The equation we need to solve is then given by

\begin{equation}
    \frac{da_{2}}{dz} \approx - \frac{\alpha_{2}}{2}a_{2} -i\kappa_{\text{eff}}\hspace{0.05cm}\text{sign}\left[\cos{\left(\Delta \beta z\right)}\right]\left[a_{1}(0)\right]^{2}e^{-\alpha_{1}z}e^{-i\Delta \beta z}.
\end{equation}

The next step is to expand the function $\text{sign}\left[\cos{\left(\Delta \beta z\right)}\right]$ into its Fourier series, which is given by

\begin{equation}
    \text{sign}\left[\cos{\left(\Delta \beta z\right)}\right] = \frac{4}{\pi}\sum_{n=0}^{\infty}\left(-1\right)^{n}\frac{\cos\left[\left(2n+1\right)\Delta \beta z\right]}{2n+1} \approx \frac{4}{\pi}\cos\left(\Delta \beta z\right).
\end{equation}

Substituting the function with only the first term of its Fourier series, rewriting and averaging out the fast term, we then have

\begin{equation}
\begin{split}
\frac{da_{2}}{dz} &\approx -\frac{\alpha_{2}}{2}a_{2} - i\frac{4}{\pi}\kappa_{\text{eff}}[a_{1}(0)]^{2}e^{-\alpha_{1}z}\cos(\Delta\beta z)e^{-i\Delta\beta z} \\
&= -\frac{\alpha_{2}}{2}a_{2} - i\frac{4}{\pi}\kappa_{\text{eff}}[a_{1}(0)]^{2}e^{-\alpha_{1}z}\frac{1+e^{-2i\Delta\beta z}}{2} \\
&\approx -\frac{\alpha_{2}}{2}a_{2} - i\frac{4}{\pi}\kappa_{\text{eff}}[a_{1}(0)]^{2}e^{-\alpha_{1}z}\cdot\frac{1}{2} \quad (\text{averaging out } e^{-2i\Delta\beta z}) \\
&= -\frac{\alpha_{2}}{2}a_{2} - i\frac{2}{\pi}\kappa_{\text{eff}}[a_{1}(0)]^{2}e^{-\alpha_{1}z}\,.
\end{split}
\end{equation}

and this equation has an analytical solution given by

\begin{equation}
    a_{2}(z) = i\frac{2}{\pi}\kappa_{\text{eff}}\left|a_{1}(0)\right|^{2}e^{-\frac{\alpha_{2}}{2}z}\left(\frac{1 - e^{\left(\frac{\alpha_{2}}{2} - \alpha_{1}\right)z}}{\frac{\alpha_{2}}{2} - \alpha_{1}}\right)
\end{equation}

Then, the absolute conversion efficiency can be written as:

\begin{equation}
    \eta_{\text{abs}} := \left|\frac{a_{2}(L)}{a_{1}(0)}\right|^{2} = \frac{4}{\pi^{2}}\left|\kappa_{\text{eff}}\right|^{2}\left|a_{1}(0)\right|^{2}L^{2}e^{-\left(\alpha_{1}+\frac{\alpha_{2}}{2}\right)L}\left(\frac{\sinh^{2}\left[\left(\alpha_{1} - \frac{\alpha_{2}}{2}\right)\frac{L}{2}\right]}{\left[\left(\alpha_{1} - \frac{\alpha_{2}}{2}\right)\frac{L}{2}\right]^{2}}\right).
\end{equation}

Even though this derivation assumed only the  $d_{zzz}$ coefficient, we emphasize that in the COMSOL simulation, all coefficients were considered non-zero and a full vectorial simulation was performed. Normalizing the fields such that the incident pump power is $\left|a_{1}(0)\right|^{2} = 1$ [W], the numerical efficiency value was found to be

\begin{equation}\label{eq:eff_comsol_full}
    \frac{\eta_{\text{abs}}}{\left|a_{1}(0)\right|^{2}L^{2}} = 9048\left[\frac{\%}{\text{W}.\text{cm}^{2}}\right]
\end{equation}

in which we used $d_{33} = 25$ [pm/V], $d_{31} = 4.9$[pm/V] and $d_{22} = 2.2$[pm/V] at 780 [nm]. The geometry parameters were $ L = 1.5$ [cm], $\alpha_{1} = 12.3$ [1/m], $\alpha_{2} = 53.0$ [1/m], waveguide top width 1750 [nm], LN thickness 600 [nm], sidewall angle 62.5$^{\circ}$ and slab thickness 80 [nm] on top of a silicon dioxide substrate. We note that  if we considered only $d_{33}$ to be non-zero, the efficiency  drops to 9047 [\%/W.cm$^{2}$]. This is only  1[\%/W.cm$^{2}$] smaller than the efficiency predicted using all components of the $d_{iJ}$ tensor, proving that indeed it is sufficient to consider only $d_{33}$ - the leading coefficient in the conversion. If we set the pump to be lossless ($\alpha_{1} = 0$) then we get an efficiency of 11084 [\%/W.cm$^{2}$]. This value increases  to  16275 [\%/W.cm$^{2}$] if both fields are lossless ($\alpha_{1} = \alpha_{2} = 0$). This supports the necessity of including linear propagation losses to properly estimate the conversion efficiency. 

\newpage
\subsection*{Ordinary differential equation (ODE) model}
The coupled ordinary differential equations for the fundamental harmonic amplitude $a_1$ and second harmonic field amplitude $a_2$ accounting for second-harmonic generation and linear absorption are given by

\begin{equation}
    \label{eq:simplifiedCMEA1}
    \frac{da_1}{dz} = - i\kappa a_2a^{*}_{1} e^{i\Delta \beta z} -\frac{\alpha_{1}}{2}a_1
\end{equation}

\begin{equation}
    \label{eq:simplifiedCMEA2}
    \frac{da_2}{dz} = - i\kappa a^2_1 e^{-i\Delta\beta z} -\frac{\alpha_{2}}{2}a_2
\end{equation}\\
where $\kappa $ is the second-harmonic generation coupling between $a_1$ and $a_2$, $z$ is the coordinate in the propagation direction, and $\Delta \beta$ is the phase mismatch. The equations are symmetrically coupled when normalized to intensity due to the degeneracy of the upconversion process $2\omega = \omega + \omega$ and the non-degeneracy of the simultaneous downconversion process $\omega = 2\omega - \omega$. Here, $\kappa $ can be derived from the experimentally obtained power and length normalized conversion efficiency $\eta$ with units [$\mathrm{\% W^{-1} cm^{-2}}$] as $\kappa = \sqrt{\eta}$. The experimentally obtained values of $\alpha_{i}$ are measured as a function of intensity, hence the factor $\frac{1}{2}$ for the field amplitude attenuation in Eqs. \ref{eq:simplifiedCMEA1} and \ref{eq:simplifiedCMEA2}. More details about our spiral loss measurements can be found in the main text and a previous section of the Supplementary Information.

So far, this model does not account for two-photon absorption (TPA), an intensity-dependent loss. The differential equation describing \textit{only} TPA is

\begin{equation}
    \frac{d I}{dz} = -\beta_{TPA} I^2
\end{equation}\\

where $\beta_{TPA}$ is the TPA coupling coefficient and $I$ is the intensity of the field \cite{Beyer2005OptLett}. Recasting this equation in terms of field amplitudes $a = \sqrt{I A_\mathrm{eff}}$, where $A_\mathrm{eff}$ is the effective area of the mode:

\begin{equation}
    \frac{d}{dz} \frac{\left|a\right|^2}{A_\mathrm{eff}} = -\beta_{TPA}\frac{\left|a\right|^4}{A_\mathrm{eff}^2}
\end{equation}

which in terms of $d|a|/dz$ is

\begin{equation}
    \frac{d\left|a\right|}{dz} = \frac{-\beta_{TPA}}{2 A_\mathrm{eff}}\left|a\right|^2 a
\end{equation}

This is the form of the TPA coupling term for a given field $a$. The TPA components of the differential equations for our two fields $a_1$ and $a_2$ are then

\begin{equation}
    \frac{da_1}{dz} = -\frac{1}{2}\left(\frac{\beta_{12}}{A_\mathrm{eff,mix}}\left|a_2\right|^2+\frac{\beta_{11}}{A_\mathrm{eff,1}}\left|a_1\right|^2\right)a_1
\end{equation}
\begin{equation}
    \frac{da_2}{dz} = -\frac{1}{2}\left(\frac{\beta_{21}}{A_\mathrm{eff,mix}}\left|a_1\right|^2+\frac{\beta_{22}}{A_\mathrm{eff,2}}\left|a_2\right|^2\right)a_2
\end{equation}\\

where $\beta_{ij}$ is the two photon absorption coefficient for photons from $a_i$ and $a_j$ and $A_\mathrm{eff,mix}$ is the average of the effective mode areas of $a_1$ and $a_2$. The energy of two second harmonic 390 nm photons exceeds the band gap of lithium niobate. So does the energy of one fundamental harmonic 780 nm photon and one second harmonic 390 nm photon - thus the cross terms $\beta_{12}$ and $\beta_{21}$.

Therefore, the complete differential equations used to model frequency conversion and linear and nonlinear losses are
\begin{equation}
    \frac{da_1}{dz} = -i\kappa a_2 a_1^* e^{i\Delta\beta z}-\frac{\alpha_{1}}{2}a_1-\frac{1}{2}\left(\frac{\beta_{11}}{A_\mathrm{eff,1}}\left|a_1\right|^2+\frac{\beta_{12}}{A_\mathrm{eff,mix}}\left|a_2\right|^2\right)a_1
\end{equation}
\begin{equation}
    \frac{da_2}{dz} = -i\kappa \left|a_1\right|^2 e^{-i\Delta\beta z}-\frac{\alpha_{2}}{2}a_2-\frac{1}{2}\left(\frac{\beta_{21}}{A_\mathrm{eff,mix}}\left|a_1\right|^2+\frac{\beta_{22}}{A_\mathrm{eff,2}}\left|a_2\right|^2\right)a_2
\end{equation}

\begin{figure}[h]
    \centering
    \makebox[\textwidth][c]{\includegraphics[width=60mm]{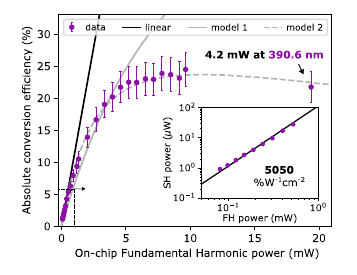}}
    \caption{\textit{Experimental data and results from the numerical model described in this section. Model 1 and 2 represent the same model solved with different parameters, as described in table \ref{tab:1} and \ref{tab:2}, respectively.}}
    \label{fig:ODEmodel1and2}
\end{figure}

A comparison of our experimental results and the theoretical model described here is shown in Fig. \ref{fig:ODEmodel1and2}. Here, model 1 refers to the model parameters based on literature values  \cite{Beyer2005PRE,Beyer2005OptLett}, as described in the main text. The parameters are allowed to vary within the uncertainties reported in their respective references. Model 2 refers to a free fit of all TPA parameters. For both models the parameters are shown in in table \ref{tab:1} and \ref{tab:2}. The fitting parameters $\alpha_{1}$ and $\alpha_{2}$ are only fit in the bounds of their experimental errors in both models. The parameters $\beta_{11}$, $\beta_{12}$, $\beta_{21}$ and $\beta_{22}$ are fit freely in model 2. The initial value for $\kappa$ is set by calculating its value from the undepleted regime of the measured data; it is then allowed to vary $\pm50$\% in both models. A fit condition $\beta_{12} = \beta_{21}$ is added for both models.

\begin{table}[h]
\centering
\caption{Model 1 parameter values}
\label{tab:1}
\begin{tabular}{llc}
\toprule
\textbf{Name} & \textbf{Value} & \textbf{Allowed to vary?} \\
\midrule

$\kappa $ & 7.662 $\mathrm{W}^{-1/2}\mathrm{cm}^{-1}$ & $\pm50$\% \\
$\alpha_{1}$ & 0.1566 $\mathrm{cm}^{-1}$ & Within uncertainty \\ 
$\alpha_{2}$ & 0.5412 $\mathrm{cm}^{-1}$ & Within uncertainty \\ 
$\beta_{11}$ & 0 cm/W & No \\
$\beta_{12}$ & 1.012e-09 cm/W & Within uncertainty \\
$\beta_{21}$ & 1.012e-09 cm/W & Within uncertainty \\
$\beta_{22}$ & 4.500e-09 cm/W & Within uncertainty \\
\bottomrule
\end{tabular}
\end{table}

\begin{table}[h]
\centering
\caption{Model 2 parameter values}
\label{tab:2}
\begin{tabular}{llc}
\toprule
\textbf{Name} & \textbf{Value} & \textbf{Allowed to vary?} \\
\midrule

$\kappa $ & 9.463 $\mathrm{W}^{-1/2}\mathrm{cm}^{-1}$ & $\pm50$\% \\
$\alpha_{1}$ & 0.1037 $\mathrm{cm}^{-1}$ & Within uncertainty \\ 
$\alpha_{2}$ & 0.5200 $\mathrm{cm}^{-1}$ & Within uncertainty \\ 
$\beta_{11}$ & 8.105e-07 cm/W & Freely \\
$\beta_{12}$ & 6.644e-09 cm/W & Freely \\
$\beta_{21}$ & 6.644e-09 cm/W & Freely \\
$\beta_{22}$ & 2.349e-09 cm/W & Freely \\
\bottomrule
\end{tabular}
\end{table}

Of particular note is the nonzero best-fit value for $\beta_{11}$ in model 2. This appears to indicate that out-of-model nonlinear loss in the fundamental harmonic field can account for our measured data. Further study of the material properties of TFLN at these wavelengths and powers is needed to understand these dynamics and the processes contributing to them.

\clearpage

\subsection*{Poling of a shallower etched waveguide}
\begin{figure}[h!]
    \centering
    \makebox[\textwidth][c]{\includegraphics[width=84.446mm]{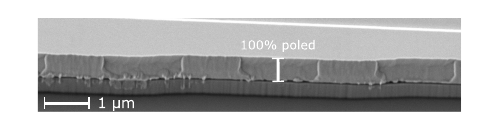}}
    \caption{\textit{Scanning electron microscope image of a sidewall poled TFLN waveguide with the inverted domains revealed by an SC-1 etch. To make this waveguide the 600 nm film is etched 300 nm instead of the 520 nm etch depth used for the main text results. At a poling voltage exceeding 250 V complete domain inversion of the film is observed.}}
    \label{fig:shallowetch}
\end{figure}

The main advantage of our \textit{pole-after-etch} method based on sidewall poling is that it results in complete (100\%) poling of the entire waveguide cross-section, resulting in a high nonlinear conversion efficiency. For these results, shown in the main text, a 600 nm lithium niobate film was etched 520 nm. We also applied our approach to  a 300 nm etched waveguide on a 600 nm lithium niobate film. After fabricating and poling the waveguide, the inverted domains are revealed using an SC1 etch (same process as discussed in the main text). From a scanning electron microscope image, Fig. \ref{fig:shallowetch}, we infer that full poling of the ridge can be achieved albeit at a higher voltage of 250 V. During earlier studies at these poling periods and with a 520 nm etch depth, these devices required about 150 V for full poling of the film. The poling period shown, approximately 3 $\upmu$m, is tailored for second harmonic generation from infra-red to visible wavelengths.

\clearpage

\newpage
\subsection*{Early exploration of narrow UV SPLN waveguides}
\begin{figure}[h!]
    \centering
    \makebox[\textwidth][c]{\includegraphics[width=180mm]{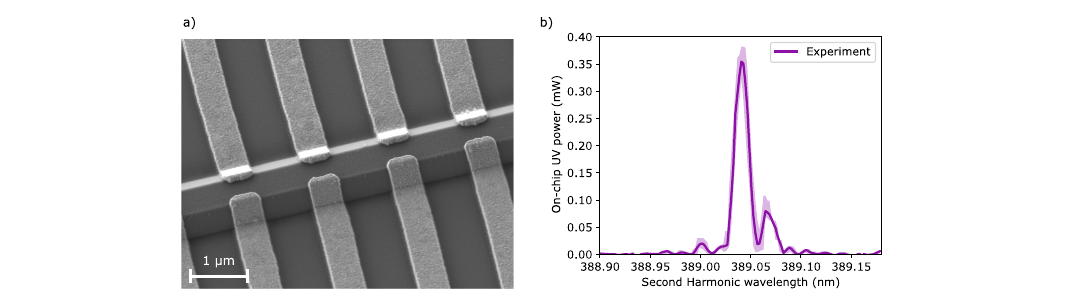}}
    \caption{\textit{a) Scanning electron microscope image of a narrower SPLN waveguide with a top width of only 0.55 $\mu$m. b) Phase matching function measured for a 0.55 $\mu$m wide and 1.5 cm long waveguide. The shaded area surrounding the solid lines denotes the experimental deviation in UV signal.}}
    \label{fig:narrowwg}
\end{figure}

Waveguides with 0.55 $\upmu$m wide top width were fabricated (Fig. \ref{fig:narrowwg}a) to explore whether a higher intensity mode waveguide would result in higher conversion efficiency. The phase matching function for a 1.5 cm long device is measured (Fig. \ref{fig:narrowwg}) and follows a single peak, $\mathrm{sinc{}^2}$ shape as expected from theory. We measure maximum on-chip UV power at 389 nm of 0.38 mW at 9.09 mW of pump power. This results in an absolute efficiency of 4.2\%, about 6x lower when compared to the 1.75 $\upmu$m wide waveguides reported in the main text. This is mainly attributed to the increased propagation losses measured for these narrower waveguides, 5.4 dB/cm vs 2.4 dB/cm.

\newpage
\printbibliography

\end{document}